\documentclass[%
    letterpaper,%
    10pt,%
    conference%
]{ieeeconf}

\IEEEoverridecommandlockouts                              

\usepackage{graphics}
\usepackage{amsmath}
\usepackage{amssymb}
\usepackage{color}
\usepackage{soul}
\usepackage{booktabs}
\usepackage{graphicx}

\let\labelindent\relax
\usepackage{enumitem}

\usepackage{mathabx}
\usepackage[hidelinks]{hyperref}
\usepackage[nameinlink]{cleveref}

\hypersetup{
  colorlinks = true, 
  urlcolor = blue, 
  linkcolor = blue, 
  citecolor = red 
}

\newcommand{\blind}[2]{\ifdefined\blinded#2\else#1\fi}

\newcommand{\ie}{\textit{i.e.}}             
\newcommand{\eg}{\textit{e.g.}}             
\newcommand{\vs}{\textit{versus}}           
\newcommand{\etc}{\textit{etc}.}           

\title{%
    \LARGE%
    \bf%
    Sociotechnical Specification for the Broader Impacts of Autonomous Vehicles$^{*}$%
}

\author{%
    \blind{%
        Thomas Krendl Gilbert$^{1}$, Aaron J. Snoswell$^{2}$, Michael Dennis$^{3}$, Rowan McAllister$^{3}$, and Cathy Wu$^{4}$%
        \thanks{%
            *This research was partly supported by the Australian Research Council Centre of Excellence for Automated Decision-Making and Society (CE200100005)%
        }%
        \thanks{%
            $^{1}$Digital Life Initiative, Cornell Tech, New York, NY, USA
            {\tt\small tg299@cornell.edu.au}%
        }%
        \thanks{%
            $^{2}$Centre for Automated Decision-Making and Society, Queensland University of Technology, Brisbane, QLD, Australia
            {\tt\small a.snoswell@qut.edu.au}%
        }%
        \thanks{%
            $^{3}$University of California, Berkeley, CA, USA
            {\tt\small [michael\_dennis|rmcallister]@berkeley.edu}%
        }%
        \thanks{%
            $^{4}$The Laboratory for Information \& Decision Systems;  the Institute for Data, Systems, and Society; and the Department of Civil and Environmental Engineering, Massachusetts Institute of Technology, Cambridge, MA, USA
            {\tt\small cathywu@mit.edu}%
        }%
    }{%
        Anonymous author(s)$^{1}$%
        \thanks{%
            $^{*}$Funding acknowledgement removed for blind peer review
        }%
        \thanks{%
            $^{1}$Affiliation(s) removed for blind peer review
        }%
    }
}

\let\svthefootnote\thefootnote

\begin{document}

\maketitle
\thispagestyle{empty}
\pagestyle{empty}

\begin{abstract}%
    Autonomous Vehicles (AVs) will have a transformative impact on society. Beyond the local safety and efficiency of individual vehicles, these effects will also change how people interact with the entire transportation system.
    This will generate a diverse range of large and foreseeable effects on social outcomes, as well as how those outcomes are distributed.
    However, the ability to control both the individual behavior of AVs and the overall flow of traffic also provides new affordances that permit AVs to control these effects.
    This comprises a problem of \textit{sociotechnical specification}: the need to distinguish which essential features of the transportation system are in or out of scope for AV development. 
    We present this problem space in terms of technical, sociotechnical, and social problems, and illustrate examples of each for the transport system components of social mobility, public infrastructure, and environmental impacts.
    The resulting research methodology sketches a path for developers to incorporate and evaluate more transportation system features within AV system components over time.
\end{abstract}


\let\thefootnote\relax\footnote{%
    \vspace{-5pt}
    \hrule
    \vspace{5pt}
    \noindent
    Paper accepted for presentation at ICRA'22 workshop ``\href{https://www.icra2022av.org/}{Fresh  Perspectives  on  the  Future  of  Autonomous  Driving}''
}
\addtocounter{footnote}{-1}\let\thefootnote\svthefootnote

\section{INTRODUCTION}
\label{sec:intro}

As AV fleets grow in size and area of service, they will comprise routing platforms with the potential to control traffic on public roads.
Such coordinated but distributed action will be unprecedented in scale, and will bring myriad new mobility challenges such as routing-consolidated infrastructure wear \cite{collier2018disrupting} and mixed-autonomy traffic jams with complicated dynamics \cite{wu2017emergent}.
There will also be new opportunities to address existing challenges such as food deserts \cite{walker2010disparities} and equitable mobility load-balancing \cite{kleinberg1999fairness}.
Moreover, growing automation will see concomitant changes as other societal institutions adapt – for instance, new mobility options and markets will supplant previous ones, changing the relative accessibility of locations, in turn impacting property markets and cost of living.
To address these dynamic trends, designers must contend with the feedback loops generated and exacerbated by coordinated optimization efforts \cite{gilbert2022choices}. Because AVs will inevitably remake the transportation system, our view is that these changes ought to be intentionally designed.

Rather than simply automating trips, a more helpful framing may be that AVs will automate core components of the transportation system itself. People will take more and different trips, and every aspect of society that interacts with the transportation system will co-adapt to take advantage of its strengths and compensate for its weaknesses.
As stakeholders in this transition, designers must help ensure that the automated transportation systems of the near future serve as intentional and active means of reshaping societal institutions. 
Recent proposals to develop \textit{world models} show the potential for training policies based on features extracted from simulation rather than from how humans drive \cite{ha2018world}.
Meanwhile, results using the Project Flow toolkit have demonstrated the potential for distributed control over mixed autonomy traffic, using existing simulation and optimization methods and models \cite{wu2017flow,wu2018stabilizing,wu2021flow}.
These examples illustrate the nascent potential of applying contemporary AI \& robotics methods to build AVs which support the automated transportation systems we \textit{want} to exist.
However, fully realising this vision will require a novel conceptual framing of the computational toolkit of contemporary AV researchers.
In this paper, we propose the frame of \textit{sociotechnical specification} as a way to understand the broader impacts of AVs on transportation systems, and give examples of ways that AV designers might begin to contend with these issues.

The remainder of this paper is structured as follows.
In \Cref{sec:ss-transit}, we motivate our framing of this problem, describing the space of problems that define what is able to be in- or out-of-scope for AV developers.
This framing can be broken down into problems that are ``technical'', ``sociotechnical'', and ``social'' in nature.
In \Cref{sec:examples}, we then give examples of transportation system components and associated features that fall into these categories, which are further elaborated in the Appendix.
\Cref{sec:conclusion} concludes.

\section{SOCIOTECHNICAL SPECIFICATION IN THE DESIGN OF `TRANSPORTATION'}
\label{sec:ss-transit}

Recent research with emerging simulation toolkits such as Project Flow illustrate that the main roadblocks to the specification of dynamic problems like mixed-autonomy traffic may no longer be strictly technical \cite{abbink2020artificial}.
Instead, there is a growing need to interrelate a transportation system's \textit{features} (i.e. its outstanding attributes or integral properties, like multi-modal access) with an AV fleet's \textit{components} (i.e. the technical capacities needed for optimal performance, like routing or sensing).
This suggests a new research agenda concerned with redesigning key features of the transportation system, rather than replicating current systems' limitations.

We argue here that translating features of transportation systems into technical components constitutes a problem of \textit{sociotechnical specification}: the need for designers to differentiate which features lie in or out of scope for AV development.
We argue that many unresolved but tractable problems related to traffic dynamics naturally follow from this shift in frame.
In particular, sociotechnical specification comprises three elements.
First, a much greater variety of features may be placed `in-scope' for AV development and performance optimization.
This moves beyond local control, perception, and route planning and includes both infrastructural features of the transportation system and those of adjacent systems (for instance environmental, economic, or municipal).
Second, AV design requires reflection about which of these features bear on a given AV component.
Particular considerations about the planning horizon, vehicle sensors, route optimization, and consumer incentives must be formalized with reference to the intended purpose of the transportation system and the role of AVs within it.
Third, more clear definitions of features will be needed for effective means of control over those that lack sufficient documentation by social scientists or law and public policy experts.
In cases where features are objects of intervention by a given AV component, designers must identify criteria for their means of control from affected systems or those experts who oversee them.

This problem framing is sociotechnical because it recognizes the possibility of translating features across abstraction scales and societal systems into elements of AV design.
As such, the three elements of sociotechnical specification outlined above suggest three corresponding problem stages:

\begin{enumerate}[leftmargin=0.5cm]

    \item \textit{Technical problems} are in need of a ``shovel'': they are more or less defined already in the academic literature, data is available and sufficiently accurate, and metrics are clear and measurable.
    The problem can readily be translated into technical methodologies and analyzed/optimized. 
    There may be remaining challenges, e.g. reward sparsity, multi-agent dynamics, nonstationarity, scalability, long horizons, etc., but they are technical in nature.
    \\
    \item \textit{Sociotechnical problems} are in need of elaboration: they have a partial technical specification, but some pieces are missing. 
    Data may be unavailable, metrics may be unclear, models may be insufficiently accurate, etc. 
    At this ``stage'', to refine the problem into one that can be rigorously analyzed, there needs to be some back-and-forth between technical and domain / social science researchers. 
    This can serve to conduct "feature selection," in which facets of the problem can be prioritized according to technical limitations (e.g. data availability) and the domain question.
    \\
    \item \textit{Social problems} are in need of further definition: understanding the feature(s) themselves is the primary challenge, as indeterminacy must be resolved before any means of control can be evaluated.
    This is at the stage of defining metrics, possibly without tight consideration of technical tools and limitations.
    There is a need for more basic social investigation, including community consultation, guidance from regulatory agencies, or research by social scientists / domain experts, rather than AV researchers.

\end{enumerate}

The modes, standards, and evaluation methods of transportation systems have developed over centuries.
New modes of road, rail, aviation, aerospace, and maritime transport have added to the variety of means of travel.
Beyond safety, transportation paths have accrued standards for throughput and environmental friendliness.
In addition, the performance of particular transportation systems are now routinely evaluated for reliability, resilience, and support for higher quality of life rather than strict measures of efficiency \cite{kaewunruen2016grand}.

As a major new transportation technology, AV fleets stand to redefine the component features of transportation systems, including the modes they comprise, the values at stake in their specification, and means of evaluation.
In particular, AVs blur the distinction between infrastructural, vehicular, and intermodal features of the road environment.
While local features are able to be better controlled, global features become more suited to observation and prediction.
Likewise, route controllers can be implemented at the micro-level (e.g.
four-way stops), meso-level (highway interchanges), or macro-level (city grids, interstate expressways), which will each require distinct value considerations.

\begin{figure}
    %
    %
    \includegraphics[width=\linewidth]{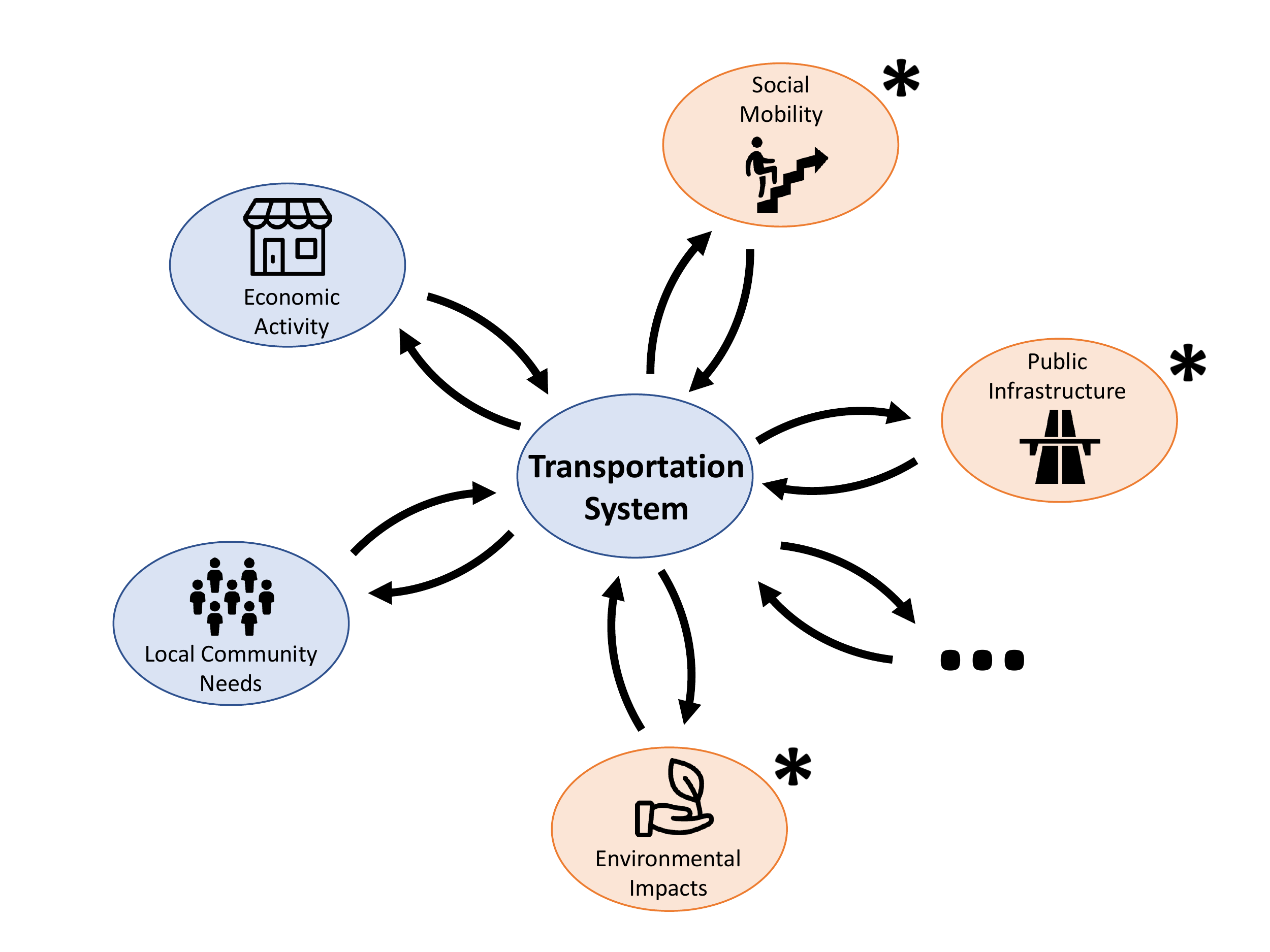}
    \vspace*{-8mm}
    \caption{%
        Transportation systems interact with numerous component features, some of which are illustrated here.
        In this article, we focus on (a) social mobility, (b) environmental impacts, and (c) public infrastructure (which are highlighted with $\Asterisk$).
    }
    \label{fig:transit-system}
\end{figure}

In this paper we focus on three component features of a transportation system, each of which stand to be significantly transformed by AV fleet operations \Cref{fig:transit-system}.
These are: (1) social mobility, (2) public infrastructure, and (3) environmental impacts.
In each setting, we give an example of a technical problem, including an example proxy metric, as well as outlining features that comprise sociotechnical and social problems.
The identification and labelling of transport system components is itself a political activity, however we have chosen these components because they are most clearly representative of the reflection that underlies sociotechnical specification.
The discussion that follows is intended illustrate this reflection and offer avenues for future refinement and elaboration.
Our hope is that future research will build on these starting points, further developing sound mathematical metrics that can serve as to highlight and control for the salient broader impacts of AVs.

\section{EXAMPLES OF TRANSPORTATION SYSTEM FEATURES AND COMPONENTS}
\label{sec:examples}

\subsection{Social Mobility}
\label{subsec:socmob}

\paragraph*{Technical problem -- food deserts}

AVs will affect access to resources for individual well-being such as food, jobs, schools, and entertainment.
As AV routing algorithms unevenly shape how easy it is to get from one location to another, they have the ability to connect —or further isolate— individuals from the resources they need to live a healthy and fulfilling life.

An example of a concrete problem that can be measured with proxy metrics, we can consider the problem of food deserts \cite{beaulac2009peer}.
Suppose we have some population living in housing geographically distributed according to $\tilde{H}$.
There are ‘good’ food options distributed by $\tilde{F}^G$ and bad food options distributed by $\tilde{F}^B$.
For a routing or AV control algorithm $A$ we define the time to go from a housing location $h$ to a food option $f$ as $t^{(A)}(h, f)$ We can then construct a simplistic consumer choice model in which a consumer at housing location $h$ will visit a high quality food option if
$
\min_{%
    (f^G,f^B) \in \tilde{F}^G \times \tilde{F}^B
}
\{
t^{(A)}(h, f^G) - t^{(A)}(h, f^B)
\}
< k,
$
for some positive constant $k$.
In other words, the consumer will choose a high-quality food option if the added cost over the easiest low-quality food option is less than $k$.
If we define $C^{(A)}(h) = 1$ if a consumer at location $h$ would choose the high quality option under algorithm $A$ and 0 otherwise, we obtain a proxy metric for consumer population nutritional health:
$
    \eta^{(A)}
    (\tilde{H})
    \triangleq
    \mathbb{E}_{h \sim \tilde{H}}
    \left[
        C^{(A)}(h)
    \right].
$

Note that this is a clearly simplified model for consumer choice, as well as the differential impact of foods on consumer health.
Nevertheless, this simple example illustrates the fact that an AV designer could in theory begin investigating the impacts of algorithmic and modeling choices on this important societal issue today.
The requisite data to inform this model (\ie{} housing and food option distributions) could be readily procured, even if in approximate forms, and existing simulation tools could then be brought to bear to analyze how choices within the remit of the AV operator or designer mediate the issue of food deserts and consumer nutritional health.

\begin{table}[t]
    \caption{Examples of transportation system features and components}
    \label{tab:menu}
    \footnotesize
    \def\arraystretch{1.5}
    \setlength{\tabcolsep}{2pt}
    \begin{tabular}{p{24mm} p{16mm} p{20mm} p{20mm}}
        \toprule
        &
        Technical
        &
        Sociotechnical
        &
        Social
        \\
        \midrule
        Social mobility
        &
        Food deserts
        &
        Housing markets
        &
        Quality of life
        \\
        Public infrastructure
        &
        Road wear
        &
        Traffic laws
        &
        Privatization of roads
        \\
        Environmental impacts
        &
        Air quality
        &
        Induced demand
        &
        Cobalt extraction
        \\
        \bottomrule
        \\
    \end{tabular}
    \vspace{-25pt}
\end{table}

\paragraph*{Sociotechnical problem -- housing markets}

Over the long-term, AV services cannot take where people happen to live as a given.
People will relocate, either buying new homes or renting temporary housing or otherwise changing their place of residence.
Where people decide to live depends on a variety of considerations such as distance from work, location of family, childcare services, or personal lifestyle, however a common denominator is the transport system’s capacity to accommodate regular use by local residents.
Natural questions arise from this: what are the aggregate effects of AV service on the dynamics of the housing market? And how might designers ensure these effects are fair or equitable?

While this question cannot be easily reduced to an econometric model, it is also not entirely new.
Much as cars facilitated the growth of urban sprawl and suburbanization \cite{norton2011fighting}, regular AV services will likely reshape both the features of neighborhoods and the calculus of how people make use of the mobility they permit.
Even once public trust in AV safety performance is assured, these effects will depend heavily on their routing algorithms, which will determine how easy it is to get from place to place and will modify the value of time of being in transit.

The criteria for specifying the learning parameters of these algorithms 
are already active research topics in adjacent literatures such as Urban Studies and Financial Economics.
However, various means of translating these topics into optimization criteria could support distinct design strategies for AV routing.
One is to scale AV infrastructure with equity-focused initiatives so that deployment does not accidentally cause gentrification or urban decay in regions where service offerings are mismatched with local mobility needs \cite{seattle}.
There may well be cross-regional consequences of this strategy, as the urban sprawl we see now is not only a result of people moving to suburbs, but the housing supply adapting to that demand.

Because use of public transit, bicycles, and pedestrian mobility increase sharply in dense environments, AVs will require routing capacities that scale with various environmental and infrastructural measures of density.
These include transit-oriented employment and residential density, development scale, density gradient, population and employment centrality, population density gradient, density at median distance, density of development, percent of houses within one mile of an elementary school, percentage increase in residential density, gross and/or net residential density, building coverage ratio, average school size, and non-residential intensity.

\paragraph*{Social problem -- Quality of Life}

The `success' or `failure' of AV adoption can be broadly understood in terms of change in overall quality of life -- the standard of health, comfort, and happiness or flourishing that is induced for and experienced by a specific group of people.
This concept, while elusive, can actually be evaluated indirectly by monitoring properties of transit and adjacent systems that have yet to receive systematic study from an organized research field.

While there are many measures of quality of life available from the field of development economics, the unprecedented dynamics induced by AV deployment require a new approach.
These dynamics often reflect deeper structural problems at the intersection of physical and economic mobility.
For example, access to the city center remains a central concern of urban planning and transportation infrastructure \cite{shen1998location}, serving as a proxy for access to labor markets and low-income mobility \cite{montgomery2018america}.
Recent measurements of food deserts \cite{beaulac2009peer}, commuter health exposure \cite{knibbs2011review}, and subjective appraisals of daily travel routes \cite{gatersleben2007affective} indicate the difficulties of tracking unanticipated externalities of common travel patterns.

A barrier to effective models in this domain is the need to identify metrics that capture both individual day-to-day decisions and longitudinal traffic effects in ways that public agencies can evaluate as good or bad.
Fortunately, there are many canonical metrics for road network usage that are also easy to communicate to the public, meaning that new models can be updated in response to stakeholder feedback.
Another strategy is to integrate AV usage with first- and last-mile mobility considerations, both to improve service in low-density areas \cite{ohnemus2016shared} and augment public transit connections so that the cost of switching between transportation modes is reduced for multi-modal road users \cite{shaheen2016mobility}.
Legacy metrics pertaining to destination accessibility should also prove useful.
Finally, translating local community priorities into perception and routing modules can draw from canonical measures of diversity in infrastructure and environment.
We suggest example metrics for these considerations in the Appendix.

\subsection{Public Infrastructure}
\label{subsec:pi}

\paragraph*{Technical problem -- road wear}

During operation, AVs interact with the existing transit infrastructure, including the signage, lane markings, and roads themselves.
Coordinated AV operations at scale could impact the deterioration of roadways.
The distribution of road deterioration will effectively be controlled by AV operators and manufacturers through the macro-level centralization of AV routing and the micro-level duplication of control schemes.
We illustrate here how a proxy metric can be constructed to capture and optimize against the first-order effects of AV operations on road deterioration.

Let $\mathcal{S}$ be the graph of road segments making up a given road network where AVs operate.
We assume that each edge $s \in \mathcal{S}$ has an associated quality rating $q(s)$ that quantifies the condition of the road along that road segment.
To capture the effect of AV operations on road deterioration, a simple model could be that the quality of a segment is reduced by some small proportion $\epsilon$ whenever it is traversed.
If the quality of a segment reaches some lower bound, then it can no longer be traversed, and must be repaired (presumably, at cost to the tax-payer).
For demonstration purposes, we could investigate the impact of having a routing algorithm account for a per-segment navigation cost proportional to the quality, serving as a proxy metric for road deterioration due to AV operations:
$
    \kappa(j, r) \triangleq  \sum_{s \in r} j \times q(s),
$
for a given route $r$ and some positive constant $j$.

Even this simplistic approach can illustrate important behaviors.
For instance, if a routing algorithm greedily minimizes the road wear navigation cost on a trip-by-trip basis, AVs will always opt to take the highest quality roadway at a given point in time, which will distribute, rather than concentrate road wear.
This model could readily be extended to incorporate more sophisticated models of road deterioration, as well as to account for variations in wear based on the type and loading of the vehicle.

\paragraph*{Sociotechnical problem -- traffic laws}

As traffic norms may evolve over time to favor optimal AV performance, organized lobbying may begin to pressure regulators to change traffic laws in ways that prioritize specific AV design considerations.
While this may well make manufacturers’ jobs easier, it would also comprise a major shift -- AVs would be disrupting the wider purpose or structure of the transit system rather than simply automating select parts.

As a result, it is reasonable for methods to consider possible changes or updates to existing traffic laws as an approach to control the impacts of AVs.
New regulations could support the addition of traffic lanes, pick-up/drop-off points, and zones where pedestrians have different rights.
Each of these interventions would allow for coordination between people and AVs, and thus it serves as an important lever of control, alongside local control, routing, and controlling stop lights.

Criteria for these considerations may be drawn from the fields of Transportation Studies and Public Policy - and we give examples of some possibilities in the Appendix.

\paragraph*{Social problem -- privatization of roads}

As long-distance platooning becomes more feasible, AVs will affect entire interstates and regions of countries in ways that transcend the jurisdiction of any single traffic authority.
Among many other effects, there is a risk that long-distance roadways will effectively be privatized for use by corporate freight operations, as other human drivers will be disincentivized from sharing the road with scores of automated trucks.
This outcome would be both unfair and unsustainable, as the costs of repairing roads would be outsourced ad hoc to a public that rarely uses them.

Designers can mitigate these risks by preparing AV modules for perception, localization, and routing that are sensitive to the public nature of roads.
This means that they would be sensitive to features of roads that are broadly inclusive of multi-modal stakeholders, and also route through them in ways that are minimally disruptive of traffic flow.
While these infrastructural features and modeling techniques remain works-in-progress, preparing for such a future will empower designers to integrate feedback from particular stakeholders as soon as mass AV platooning and routing is common.

With respect to routing, it is possible to measure and minimize loss in AV fuel efficiency or average time to destination from avoiding potholes, aiming to preserve the road without disrupting traffic.
This is analogous to other settings where AV software compensates for hardware limitations, except here the road itself is also modeled as “hardware” rather than relegated to the external environment.
This is quantified through reference to existing models for highway maintenance \cite{theberge1987development}, priority damage assessment \cite{snaith1984priority}, and smart pavement evaluations \cite{asbahan2008evaluating}.
This work could aid constraint satisfaction by including factors that corroborate existing public standards for road maintenance, rather than modeling vehicle motion in isolation.

Meanwhile, external changes to signage and lane markings could be modeled as controls on the large-scale effects of AVs.
A natural implication of this would incorporate measures that reflect different design scales for multi-modal concerns on a spectrum of local \vs{} global effects.
On the former end of the scale, updating perception to conform to various point design measures will help AVs modify their speed and behavior in real time to conform to stakeholder expectations and priorities.
For example, residential neighborhoods in the United States often accommodate special needs groups through distinct signage: warning signs about pet dogs and cats, ``children at play'', and protection for the disabled (\eg{} audible walk signs).
Additional examples may be found in the Appendix.

Each of these considerations, and other details of local customs which we have yet to consider, need to be incorporated into the local control procedure so that they can be customized to be contextually appropriate.
For example, beyond vehicle features such as wheelchair access, AVs will need to incorporate routing adjustments so that time spent in these zoned areas is minimized.
In particular, there is a need for regional design measures that are tailored to capture the completeness of regional transportation systems.
Meanwhile, some communities require unique forms of road mobility, such as retirement facilities and golf courses that have their own specialized modes of transport and which have adopted special guidelines for golf carts interacting with normal traffic vehicles \cite{head2012multi}.
And on a wider scale, network-level effects should also be reflected in metrics for connectivity and route directness at the neighborhood level.
Examples of all these features may be found in the Appendix.

\subsection{Environmental Impacts}
\label{subsec:ei}

\paragraph*{Technical problem -- air quality}

A growing problem in many urban cities is the contribution of transit-related pollution to the deterioration of air quality.
Vehicles release various forms of pollution which are harmful to the residents of high traffic areas, and there has been a large body of research in quantifying both the amount and the health effects of pollution on local residents \cite{fisher2002health,krzyzanowski2005health,lipfert2006traffic,zhang2013air,west2004distributional}.
The effects of this pollution will depend on the type of vehicle, including the power source (gas / hybrid / electric, \etc{}), the density of traffic, weather conditions, the type of neighborhood being traversed, physical proximity of road users, and many other complexities.

As a simple model for these effects, we can estimate the spatial distribution of people present in a region of AV operations and assume a small penalty per second for every person, inversely proportional to the square of the distance of that person to the vehicle.
Formally, if we have a spatial population of people $p \sim \tilde{P}$ and $v$ is the location of the AV, we can define a proxy metric for the impact of pollution as
$
    \pi(v, \tilde{P}) \triangleq \mathbb{E}_{p \sim \tilde{P}}
    \| p - v \|_2^{-2}.
$

An AV that is trying to minimize this cost would be more likely to route around local communities towards less occupied areas, distributing pollution where it has fewer adverse effects.
Additionally, even along a fixed route, AV designers could take advantage of the fine-grained control available to AVs, to optimize for pollution effects based on the characteristics of the population, the vehicle engine, and traffic conditions.

Population data to inform this model could be drawn from a number of sources such as county or council residential housing and zoning records, or approximated \eg{} according to lot size data from open-source map databases.
This simple model could be improved in a number of ways. 
For example, the distribution $\tilde{P}$ could be extended to cover temporal fluctuations in population density, or better models of health effects could be incorporated.

\paragraph*{Sociotechnical problem -- induced demand}

In the longer term, AVs will have uncertain effects on aggregate pollution arising from traffic.
Impacts of the AV fleet itself could be modeled without incorporating considerations of vehicle type and traffic density.
However, it is important to consider the effects of induced demand \cite{lee1999induced}.
For example, simulated results show that widespread deployment of AVs could slash U.S. energy consumption by as much as 40\% due to improved driving efficiency; alternatively, it could double U.S. energy consumption due to increased availability of cheap transport options \cite{wadud2016help}.

Even these results do not incorporate differences in road conditions, models of vehicle, fluctuations in demand, traffic conditions, or interactions between routing vehicles.
Still, this work lends support to carbon-pricing or surge pricing policies as AVs are more widely adopted.
Extending this to incorporate more work on quantifying CO$_2$ emissions \cite{west2004distributional,sgouridis2011air,noland2006flow} and induced demand \cite{hymel2019if,omstedt2005model} are clear future research directions.
As such, efforts to inquire into and model AV-induced transit demand comprises an emerging subfield at the intersection of Environmental Engineering, Behavioral Economics, and Traffic Modeling.

To further illustrate this problem, we can consider the various pollution effects beyond individual car exhaust or aggregate CO$_2$ emissions induced by AVs.
These range from noise pollution \cite{campello2017effect} to the wider ecological, fiscal, and social factors associated with AVs’ environmental sustainability.
Ecological metrics have been well documented in the literature.
Changes to them will likely generate effects on fiscal metrics related to activity level.
And changes to fiscal metrics, in turn, may have community-level social impacts whose measurement is vital but somewhat more speculative for activity level and modal share.
Examples of all three metric types may be found in the Appendix.

Incorporating these factors would have several modeling benefits.
It would provide stronger estimates for demand that would help align emissions control with wider societal aims for equitable road access.
It would also help integrate AV policy development with ongoing research on updates to pavement materials and construction practices of tollways \cite{al2015scenarios}, leading to possible new improvements in sustainability.
And it could leverage vehicle-level data collection to control for the granular spatial and temporal features of air pollution that have recently been measured at unprecedented micro scales \cite{apte2017high,caubel2019distributed}, ensuring the benefits of emissions reduction are both locally and globally equitable.
The deployment of AVs at distinct scales of road infrastructure (urban core, commuter routes, interstate highways) could also aid in the evaluation of alternative measurement approaches that trade- off precision against efficiency, an open research question in environmental engineering \cite{messier2018mapping}.

\paragraph*{Social problem -- cobalt extraction}

Like all modes of transit, AV fleets will comprise the finished endpoint of a much larger supply chain.
And like all electric cars, AVs will depend heavily on the extraction of rare earth minerals from seafloor deposits and regions of the developing world.
For example, worldwide demand for cobalt has skyrocketed due to its importance in batteries of electric cars, with countries such as the Democratic Republic of Congo becoming targets of major international investment \cite{calvao2021cobalt}.
Beyond greenhouse gas emissions, AV fleets will bring the geopolitics of mineral markets into scope for development -- analogous to how 21st century trends in offshore production and increasing consumer awareness have forced consumer-facing businesses to audit their own supply chains for environmental, sustainability, and human rights concerns.

The extraction of rare earth metals requires consideration of long-term economic development, national security, and human rights.
For example, investigations have revealed that child labor is commonly used in cobalt extraction, which often involves digging for the mineral by hand \cite{sovacool2021subterranean}.
This cobalt is often refined and prepared for manufacture by foreign companies and then used to power smartphones designed and sold by tech companies in Western post-industrial countries.
However, no U.S. laws attempt to verify the prevention of human rights abuses in cobalt’s extraction \cite{kelly2019apple}.
And electric vehicle batteries typically include hundreds of times the amount of cobalt as a laptop \cite{campbell2020cobalt}.

The ethical problems with scaling AV fleets that rely on these supply chains are stark, and engineers must prioritize the experimental design of AVs based on alternative minerals whose extraction and use do not support inhumane work practices \cite{ryu2021reducing}.
Where use of such minerals is inevitable, AV manufacturers will need to evaluate how growth of a given platform could leave its supply chain dependent on these practices, and how this dependency might otherwise be mitigated over time.

Beyond ethics, there are other reasons why reliance on the extraction of rare earth metals should be avoided.
For one, it exposes AV services to geopolitical forces beyond the control of their parent firm.
For example, local corruption among government officials or labor groups could result in supply disruptions that make it prohibitive to grow or sustain the AV fleet without drastically raising costs \cite{callaway2018powering}.
Other geopolitical actors enmeshed in a military conflict with the country of extraction 
could interfere with supply lines or plunder mineral resources for themselves.

In a broader sense, these supply chains introduce uncertainties and structural risks that make long-term forecasting of AV development impossible.
Because sources of extraction lack the transparency and documentation of labor-protected working conditions, it is difficult to evaluate exactly where and how much mineral material will be extracted over time \cite{frankel2016cobalt}. 
Consider also the prospect of international sanctions against certain forms of mineral extraction, or consumer boycotts of tech services for their reliance on them \cite{white1996globalization}.
International watchdogs and human rights groups may expose work practices so reprehensible that AV companies could be exposed to liability for human rights abuses or international condemnation.

These considerations point to an uncomfortable truth: no matter how safe and efficient and consumer-friendly AVs may become, their production remains dependent on the contradictions, ambiguities, and political traps of 21st century capitalism.
AV development must remain sensitive to these supply chain issues as their deployment continues.

\section{CONCLUSION}
\label{sec:conclusion}

Sociotechnical specification remains a looming challenge for AV fleet performance. 
While the short-term priority is building AVs that successfully conform to the existing rules of the road, over the long term AVs are more and more likely to set the pace of traffic in their own right. 
Moreover, as AVs become widely deployed, their effects and impact on public infrastructure may be unequal even as their terms are hard to specify.
Nevertheless, sustained attention to technical problems (including proxy metrics), sociotechnical problems (including adjacent literatures), and social problems (including outstanding research questions) will prepare designers to incorporate stakeholder needs and requests as they arise over time.




\section*{ACKNOWLEDGMENT}

Icons in \Cref{fig:transit-system} are CC:BY licensed from NounProject.com: \textit{Highway} by Georgiana Ionescu, \textit{Crowd} by Adrien Coquet, \textit{Environment} by Shmidt Sergey, \textit{Shop} by revo250, and \textit{Improvement} by Cuputo.

\section*{APPENDIX A}
\label[appendix]{app:appendix_a}

Sociotechnical specification is directly relevant to outstanding research topics in autonomous driving design.
This appendix highlights analogous problem spaces and corresponding tools, datasets, and evaluation methods already established in interdisciplinary literatures.
Some of these standards may be directly applicable to existing techniques, while others may require iteration or refinement. We place them here as outstanding examples of technical work that could be immediately undertaken by AV designers.

\subsection{Approaches that account for interactions between traditional subcomponents (localization, perception, planning, routing)}
\begin{itemize}
    \item \textbf{New traffic laws}:
    While AV designers are immediately concerned with optimizing AV performance rather than deciding what ``good'' performance necessarily means, it would be reasonable for methods to consider possible changes or updates to existing traffic laws as an approach to control the impacts of AVs.
    For instance, new regulations could support the addition of traffic lanes, pick-up/drop-off points, and zones where pedestrians have different rights.
    Each of these interventions would allow for coordination between people and AVs, and as such it would serve as an important interface between means of local control, routing, and coordination with external road infrastructure (e.g. controlling stop lights).
    Attention to such interfaces is likely going to be a key nexus of regulatory attention in the coming decades.
    One source of inspiration is analytical tools from the Highway Safety Manual (HSM) \cite{part2010highway}, which could be applied to a wider range of urban settings beyond highways. 
    While focused on mitigating crash frequency, the Highway Safety Manual aims to coordinate safety and economic concerns in a way that well-approximates the human factors interpretation of safety as a problem of limited attention and human capabilities \cite{banihashemi2011highway}.
    The HSM could help update current AV simulation work to prepare it for future traffic laws by embedding system planning within engineering, construction, and maintenance as part of an integrated development process.
    This perspective would prepare AV designers for federal and state regulatory environments once they have moved past proprietary standards for simulation and control.
\end{itemize}

\subsection{The collection, curation, and sampling of datasets for end-to-end or learned decision making approaches}
\begin{itemize}
    \item \textbf{Transit systems engineering}:
    The HSM discussed above also pinpoints three neglected sources of data (site characteristics data, traffic volume data, crash data) and incorporates them as part of safety prediction. 
    This would help AV designers move beyond ``cookbook engineering'' when setting up simulation parameters, and instead incorporate the basic concepts of systems engineering (functions, requirements, and context) that will ensure simulations accommodate the multiple interfaces necessary for fair and inclusive urban AV navigation.
\end{itemize}
    
\subsection{Methods for validating / evaluating the performance of end-to-end / learned approaches in simulation}
\begin{itemize}
    \item \textbf{Highway capacity}:
    A major stumbling-block for the use of traffic efficiency models is the need for more sophisticated travel time metrics for highly-localized neighborhoods, urban sub-regions, and particular corridors.
    One path forward is to incorporate Highway Capacity Software in support of the highway capacity manual \cite{manual2000highway}.
    This will help permit a choice of advanced modeling tools in the context of stakeholder interests and targeted focus groups \cite{flannery2004highway}. 
    It also makes possible particular auto-oriented metrics of demand and system utilization.
    \item \textbf{Tolling}:
    There will be a need for entirely new traffic regulations, as the maturation of AV optimization interacts with legacy forms of traffic control. 
    Designers will therefore need tools and methods to accommodate this likelihood.
    A good source of inspiration is the work on standards, metrics, and simulation parameters by the International Bridge, Tunnel and Turnpike Association (IBTTA).
    IBTTA has supported and made possible studies of the impacts of innovative technologies on highway operators \cite{azmat2018impact}, as well as the impact of public-private partnerships on financing road infrastructure in developing economies \cite{queiroz2013public}.
    IBTTA has also developed specialized tools for modeling various tolling environments.
    For example, the IBTTA Tollminer is a visualization tool that includes maps of toll facilities, a list of managed lane projects in operation nationwide, an optimizable user interface, and annual data on public and private toll revenues, among other features. While it is geared towards modeling and comparing the relative effects of high-occupancy vehicle lanes and toll lanes \cite{poole2020impact}, this work could be readily translated to test new simulation parameters for AVs that incorporate speculative regulations for equitable mobility access.
\end{itemize}

\section*{APPENDIX B}
\label[appendix]{app:appendix_b}

\Cref{tab:features} shows lists a number of other tractable metrics that may be relevant for quantifying the impacts of Autonomous Vehicles on Transit Systems.

\begin{table*}[p]
    \caption{Tractable feature types and example metrics that interact with the transportation system. Terms are taken from \cite{elefteriadou2012expanded} unless otherwise noted.}
    \label{tab:features}
    \def\arraystretch{1.5}
    \begin{tabular}{p{55mm} p{60mm}  p{50mm}}
        \toprule
        Public Infrastructure
        &
        Social Mobility
        &
        Environment Impacts
        \\
        \midrule
        \vspace{2pt}
   Privatization of roads
   \begin{itemize}
       \item Vehicle perception of signage
       \begin{itemize}
           \item wayfinding information
           \item sidewalk quality/width/shade
           \item treelined/shaded streets
           \item walkable streets
           \item systematic pedestrian and cycling environmental scan instrument
           \item commercial onsite amenities to support alternative modes
           \item availability of on-site bicycle amenities
           \item pedestrian scaled lighting
           \item ratio of street width to building height
           \item parking screening
           \item bus pass program utilization
           \item parking shading
       \end{itemize}
       \item Regional transit system components
       \begin{itemize}
           \item on-vehicle bicycle-carrying facilities
           \item parkand-rides with express service
           \item parking spaces designated for carpools or vanpools
           \item transit passes
           \item traffic cells
           \item percent miles bicycle and/or pedestrian accommodations
           \item miles of express fixed-transit route/dedicated bus lanes
           \item road density
           \item lane miles per capita
           \item percent of network that is “effective”
           \item roadway network balance
           \end{itemize}
        \item Local route connectivity
        \begin{itemize}
            \item square feet of pathways/sidewalks
            \item crosswalk spacing
            \item number of safe crossings per mile
            \item bicycle parking at stops and stations
            \item parking footprint
            \item block length
            \item parking location
            \item bicycle path condition
            \item pedestrian/bicycle route directness
            \item land use buffers
            \item walking environment
            \item bicycle maintenance stations
            \item bicycle/pedestrian connectivity
            \item connectivity indexes \cite{mishra2012performance}
            \item project adjacency to existing network
            \item connected and open community
            \item connected sidewalks/paths
            \item connected streets
            \item cross access
        \end{itemize}
       \end{itemize}
        &
        \vspace{2pt}
        Quality of Life 
        \begin{itemize}
       \item Road network usage
       \begin{itemize}
           \item vehicle occupancy by land use
           \item district-wide Level of Service (LOS)/Quality of Service (QOS)
           \item local traffic diversion
           \item percent of system heavily congested
           \item v/C ratio
           \item vehicle density
           \item demand/capacity ratio
           \item Maximum Service Volume
           \item Peak Hour LOS
           \item Percent of Capacity Consumed
       \end{itemize}
       \item Last mile
       \begin{itemize}
           \item average trip length per traveler
           \item delay per traveler
           \item door to door travel time
           \item HCM-based bicycle LOS
           \item proportion of total person miles traveled (PMT) for nonsingle occupancy vehicles (SOVs)
           \end{itemize}
        \item Land use optimization
        \begin{itemize}
            \item residence proximity
            \item employment proximity
            \item work accessibility
            \item number of key destinations accessible via a connected pedestrian system
            \item industrial/warehouse proximity
            \item transit convenience/stop accessibility
            \item geographic service coverage
            \item population service coverage
            \item percent in proximity
            \item transit accessibility
        \end{itemize}
        \item Transit mode optimization
        \begin{itemize}
            \item bike/pedestrian accessibility
            \item destination accessibility
            \item residential accessibility
            \item average walking distance between land use pairs
            \item spacing between village centers
            \item multiple route choices
        \end{itemize}
        \item Environmental diversity
        \begin{itemize}
            \item Smart Growth Index
            \item significant land uses
            \item land use ratios
            \item land use balance
            \item variation of agriculture of green fields
            \item land consumption
            \item core land use
            \item land use separation
            \item Transportation-Efficient Land Use Mapping Index (TELUMI) model
            \item minimum thresholds of land use intensity
            \item nearby neighborhood assets
            \item distinct indexes for sprawl \cite{ewing2003urban,galster2001wrestling}
            \item land use within village center
            \item land use within transit supportive area
            \item jobs/housing balance
        \end{itemize}
       \end{itemize}
        \vspace{10pt}
        &
        \vspace{2pt}
        Induced demand 
        \begin{itemize}
       \item Ecological metrics
       \begin{itemize}
           \item attainment of ambient air quality standards
           \item daily CO2 emissions
           \item daily NOx/CO/Volative organic compound (VOC) emissions
           \item noise pollution
           \item impact on wildlife habitat
           \item water runoff
       \end{itemize}
       \item Fiscal costs
       \begin{itemize}
           \item additional fuel tax
           \item transportation utility fee (TUF)
           \item vehicle miles traveled (VMT)-based impact fee
           \item consumption-based mobility fee
           \item improvements-based mobility fee
           \item cost recovery from alternate sources
           \item variable fees based on LOS
           \item benefit cost ratio
           \item parking pricing
           \item capita funding for bike/pedestrians
           \end{itemize}
        \item Community-level impacts
        \begin{itemize}
            \item distribution of benefit by income group
            \item transportation affordability
            \item equitable distribution of accessibility
            \item commute cost
            \item transit values
            \item fee charged for employee parking spaces
            \item travel demand management (TDM) effectiveness based on TRIMMS model
            \item travel costs by income group and/or race
            \item VMT by income group and/or race
            \item mode share by income group and/or race
            \item walk to transit by income group and/or race
        \end{itemize}
       \end{itemize} 
        \\
        \bottomrule
        \\
    \end{tabular}
\end{table*}


\bibliographystyle{IEEEtran}
\bibliography{references}

\end{document}